\documentclass[12pt,prd,aps,amssymb,amsmath,tightenlines,showkeys]{revtex4}

\usepackage{graphicx}

\newcommand{\half}{\textstyle{\frac{1}{2}}}

\newcommand{\cP}{{\cal P}}
\newcommand{\cT}{{\cal T}}

\begin{document}

\title{Complex Elliptic Pendulum}
\author{Carl~M.~Bender${}^1$}\email{cmb@wustl.edu}
\author{Daniel~W.~Hook${}^2$}\email{d.hook@imperial.ac.uk}
\author{Karta Kooner${}^2$}\email{karta.kooner08@imperial.ac.uk}

\affiliation{${}^1$Department of Physics, Washington University, St. Louis, MO
63130, USA \\
${}^2$Theoretical Physics, Imperial College London, London SW7 2AZ, UK}

\date{\today}

\begin{abstract}
This paper briefly summarizes previous work on complex classical mechanics and
its relation to quantum mechanics. It then introduces a previously unstudied
area of research involving the complex particle trajectories associated with
elliptic potentials.
\end{abstract}


\maketitle

\section{Introduction}
\label{s1}

In generalizing the real number system to the complex number system one loses
the ordering property. (The inequality $z_1<z_2$ is meaningless if $z_1,\,z_2\in
\mathbb{C}$.) However, extending real analysis into the complex domain is
extremely useful because it makes easily accessible many of the subtle features
and concepts of real mathematics. For example, complex analysis can be used to
derive the fundamental theorem of algebra in just a few lines. (The proof of
this theorem using real analysis alone is long and difficult.) Complex analysis
explains why the Taylor series for the function $f(x)=1/\left(1+x^2\right)$
converges in the finite domain $-1<x<1$ even though $f(x)$ is
smooth for all real $x$.

This paper describes and summarizes our ongoing research program to extend
conventional classical mechanics to complex classical mechanics. In our
exploration of the nature of complex classical mechanics we have examined a 
large class of analytic potentials and have discovered some remarkable
phenomena, namely, that these systems can exhibit behavior that one would
normally expect to be displayed only by quantum-mechanical systems. In
particular, in our numerical studies we have found that complex classical
systems can exhibit tunneling-like behavior.

Among the complex potentials we have studied are periodic potentials, and we
have discovered the surprising result that such classical potentials can have
band structure. Our work on periodic potentials naturally leads us to study the
complex classical mechanics of {\it doubly}-periodic potentials. We have chosen
to examine elliptic potentials because such functions are analytic and thus have
well-defined complex continuations. This is a rich and previously unexplored
area of mathematical physics, and in the current paper we report some new
discoveries.

This paper is organized as follows: In Sec.~\ref{s2} we give a brief review of
complex classical mechanics focusing on the complex behavior of a classical
particle in a periodic potential. In Sec.~\ref{s3} we describe the motion of a
particle in an elliptic-function potential. Finally, in Sec.~\ref{s4} we make
some concluding remarks and describe the future objectives of our research
program.

\section{Previous Results on Complex Classical Mechanics}
\label{s2}

During the past decade there has been an active research program to extend
quantum mechanics into the complex domain. Specifically, it has been shown that
the requirement that a Hamiltonian be Dirac Hermitian (we say that a Hamiltonian
is {\it Dirac-Hermitian} if $H=H^\dag$, where $\dag$ represents the combined
operations of complex conjugation and matrix transposition) may be broadened to
include complex non-Dirac-Hermitian Hamiltonians that are $\cP\cT$ symmetric.
This much wider class of Hamiltonians are physically acceptable because they
possess two crucial features: (i) their eigenvalues are all real, and (ii) they
describe unitary time evolution. (We say that a Hamiltonian is $\cP\cT$
symmetric if it is invariant under combined spatial reflection and time
reversal.)

An example of a class of Hamiltonians that is not Dirac Hermitian but which
is $\cP\cT$ symmetric is given by
\begin{equation}
H=p^2+x^2(ix)^\epsilon\qquad(\epsilon>0).
\label{e1}
\end{equation}
When $\epsilon=0$, $H$, which represents the familiar quantum harmonic
oscillator, is Dirac Hermitian. While $H$ is no longer Dirac Hermitian when
$\epsilon$ increases from $0$, $H$ continues to be $\cP\cT$ symmetric, and its
eigenvalues continue to be real, positive, and discrete
\cite{R1,R2,R3,R4,R5,R6,R7,R8,R9}. Because a $\cP\cT$-symmetric quantum system
in the complex domain retains the fundamental properties required of a physical
quantum theory, much theoretical research on such systems has been published and
recent experimental observations have confirmed some theoretical predictions
\cite{R10,R11,R12,R13}.

Complex quantum mechanics has proved to be so interesting that the research
activity on $\cP\cT$ quantum mechanics has motivated studies of complex
classical mechanics. In the study of complex systems the complex as well as the
real solutions to Hamilton's differential equations of motion are considered. In
this generalization of conventional classical mechanics, classical particles are
not constrained to move along the real axis and may travel through the complex
plane.

Early work on the particle trajectories in complex classical mechanics is
reported in Refs.~\cite{R2,R14}. Subsequently, detailed studies of the complex
extensions of various one-dimensional conventional
classical-mechanical systems were undertaken: The remarkable properties of
complex classical trajectories are examined in Refs.~\cite{R15,R16,R17,R18,R19}.
Higher dimensional complex classical-mechanical systems, such as the
Lotka-Volterra equations for population dynamics and the Euler equations for
rigid body rotation are discussed in Refs.~\cite{R20}. The complex $\cP
\cT$-symmetric Korteweg-de Vries equation has also been studied
\cite{R21,R22,R23,R24,R25,R26,R27}.

The objective in extending classical mechanics into the complex domain is to
enhance our understanding of subtle mathematical and physical phenomena. For
example, it was found that some of the complicated properties of chaotic systems
become more transparent when extended into the complex domain \cite{R28}. Also,
studies of exceptional points of complex systems have revealed interesting and
potentially observable effects \cite{R29,R30}. Finally, recent work on the
complex extension of quantum probability density constitutes an advance in
our understanding of the quantum correspondence principle \cite{R31}.

\begin{figure}
\begin{center}
\includegraphics[scale=0.28, bb=0 0 1000 692]{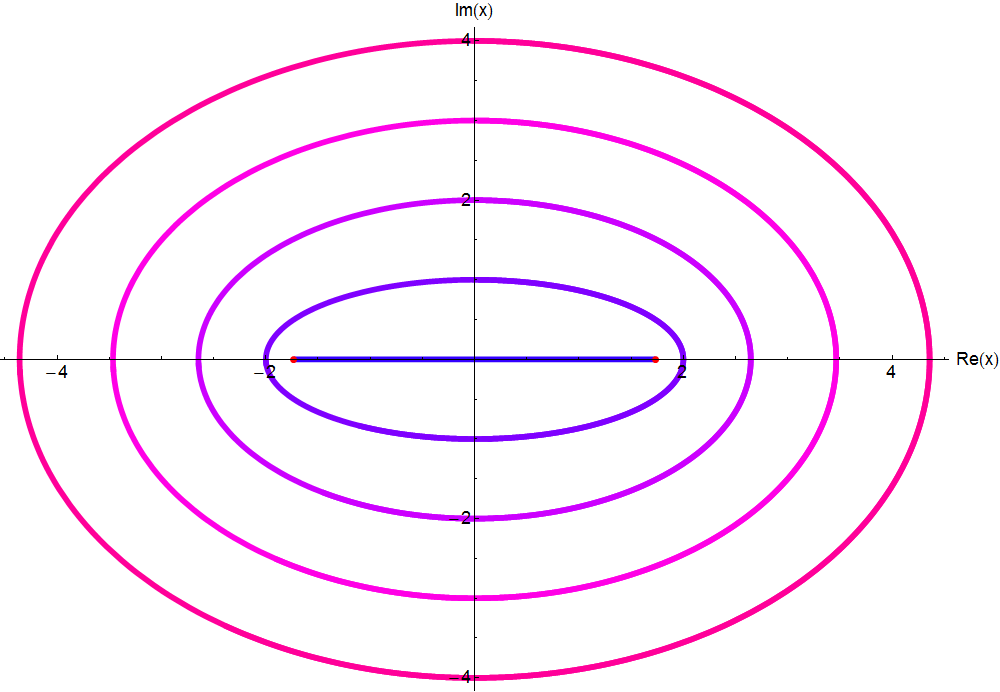}
\end{center}
\caption{Classical trajectories in the complex plane for the harmonic-oscillator
Hamiltonian $H=p^2+x^2$. These trajectories are nested ellipses. Observe that
when the harmonic oscillator is extended into the complex-$x$ domain, the
classical particles may pass through the classically forbidden regions on the
real axis outside the turning points. When the trajectories cross the real axis,
they are orthogonal to it.}
\label{F1}
\end{figure}

An elementary example that illustrates the extension of a conventional
classical-mechanical system into the complex plane is given by the classical
harmonic oscillator, whose Hamiltonian is given in (\ref{e1}) with $\epsilon=0$.
The standard classical equations of motion for this system are
\begin{equation}
\dot{x}=2p,\quad\dot{p}=-2x.
\label{e2}
\end{equation}
However, we now treat the coordinate variable $z(t)$ and the momentum variable
$p(t)$ to be {\it complex} functions of time $t$. That is, we consider this
system to have one {\it complex} degree of freedom. Thus, the equations of
motion become
\begin{equation}
\dot{r}=2u,\quad\dot{s}=2v,\quad\dot{u}=-2r,\quad\dot{v}=-2s,
\label{e3}
\end{equation}
where the complex coordinate is $x=r+is$ and the complex momentum is $p=u+iv$.
For a particle having real energy $E$ and initial position $r(0)=a>\sqrt{E}$,
$s(0)=0$, the solution to (\ref{e3}) is
\begin{equation}
r(t)=a\cos(2t),\quad s(t)=\sqrt{a^2-E}\sin(2t).
\label{e4}
\end{equation}
Thus, the possible classical trajectories are a family of ellipses parametrized
by the initial position $a$:
\begin{equation}
\frac{r^2}{a^2}+\frac{s^2}{a^2-E}=1.
\label{e5}
\end{equation}
Five of these trajectories are shown in Fig.~\ref{F1}. Each trajectory has the
same period $T=\pi$. The degenerate ellipse, whose foci are the turning points
at $x=\pm\sqrt{E}$, is the familiar real solution. Note that classical particles
may visit the real axis in the classically forbidden regions $|x|>\sqrt{E}$, but
that the elliptical trajectories are {\it orthogonal rather than parallel to the
real-$x$ axis}.

In general, when classical mechanics is extended into the complex domain,
classical particles are allowed to enter the classically forbidden region.
However, in the forbidden region there is no particle flow parallel to the real
axis and the flow of classical particles is {\it orthogonal} to the axis. This
feature is analogous to the vanishing flux of energy in the case of total
internal reflection.

In the case of total internal reflection when the angle of incidence is less
than a critical value, there is a reflected wave but no transmitted wave. The
electromagnetic field does cross the boundary and this field is attenuated
exponentially in a few wavelengths beyond the interface. Although the field does
not vanish in the classically forbidden region, there is no flux of energy; that
is, the Poynting vector vanishes in the classically forbidden region beyond the
interface. We emphasize that in the physical world the cutoff at the boundary
between the classically allowed and the classically forbidden regions is not
perfectly sharp. For example, in classical optics it is known that below the
surface of an imperfect conductor, the electromagnetic fields do not vanish
abruptly. Rather, they decay exponentially as functions of the penetration
depth. This effect is known as {\it skin depth} \cite{R32}.

\begin{figure}
\begin{center}
\includegraphics[scale=0.4, viewport=0 0 1000 605]{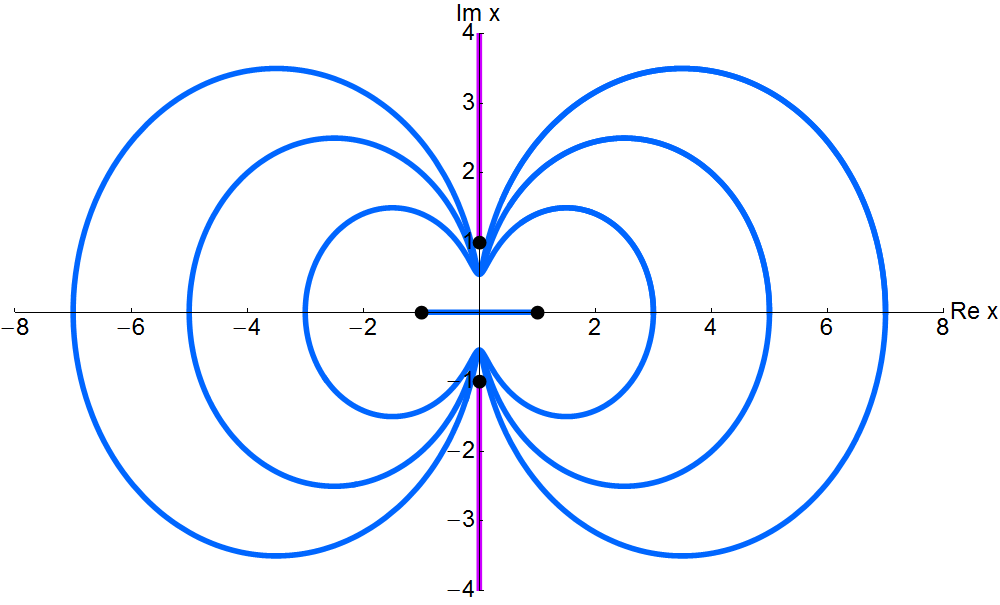}
\end{center}
\caption{Classical trajectories $x(t)$ in the complex-$x$ plane for the
anharmonic-oscillator Hamiltonian $H=\frac{1}{2}p^2+x^4$. All trajectories
represent a particle of energy $E=1$. There is one real trajectory that
oscillates between the turning points at $x=\pm1$ and an infinite family of
nested complex trajectories that enclose the real turning points but lie inside
the imaginary turning points at $\pm i$. (The turning points are indicated by
dots.) Two other trajectories begin at the imaginary turning points and drift
off to infinity along the imaginary-$x$ axis. Apart from the trajectories
beginning at $\pm i$, all trajectories are closed and periodic. All orbits in
this figure have the same period $\sqrt{\pi/2}\,\Gamma\left(\frac{1}{4}\right)/
\Gamma\left(\frac{3}{4}\right)=3.70815\ldots$.}
\label{F2}
\end{figure}

Another model that illustrates the properties of the classically allowed and
classically forbidden regions is the anharmonic oscillator, whose Hamiltonian is
\begin{equation}
H=\half p^2+x^4.
\label{e6}
\end{equation}
For this Hamiltonian there are four turning points, two on the real axis and
two on the imaginary axis. When the energy is real and positive, all the
classical trajectories are closed and periodic except for two special
trajectories that begin at the turning points on the imaginary axis. Four
trajectories for the case $E=1$ are shown in Fig.~\ref{F2}. 

The topology of the classical trajectories changes dramatically if the classical
energy is allowed to be complex: When ${\rm Im}\,E\neq0$, the classical paths no
longer closed. This feature is illustrated in Fig.~3, which shows the path of a
particle of energy $E=1+0.1i$ in an anharmonic potential.

\begin{figure}
\begin{center}
\includegraphics[scale=0.3, viewport=0 0 1000 580]{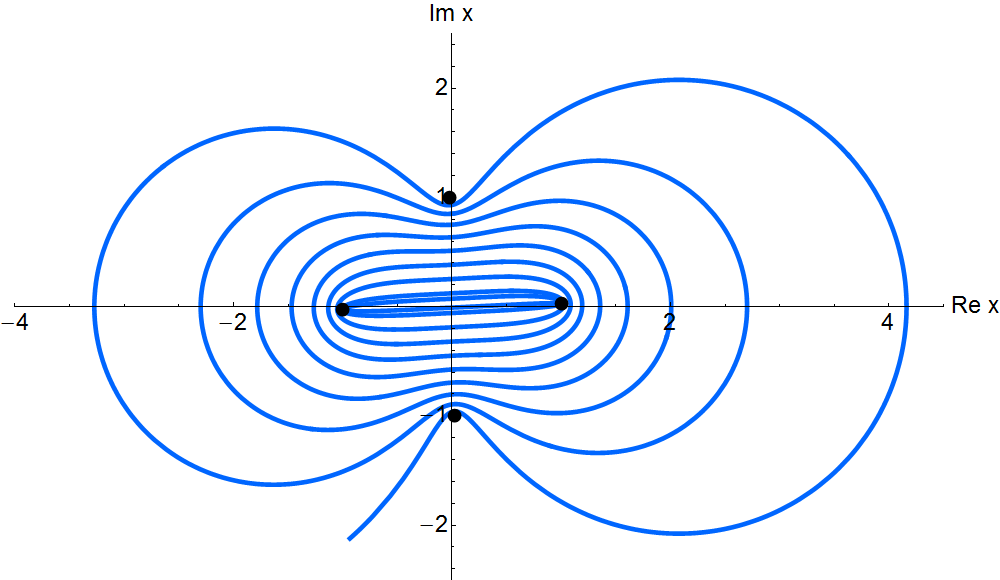}
\caption{A single classical trajectory in the complex-$x$ plane for a particle
governed by the anharmonic-oscillator Hamiltonian $H=\frac{1}{2}p^2+x^4$. This
trajectory begins at $x=1$ and represents the complex path of a particle whose
energy $E=1+0.1i$ is complex. The trajectory is not closed or periodic. The four
turning points are indicated by dots. The trajectory does not cross itself.}
\end{center}
\label{F3}
\end{figure}

The observation that classical orbits are closed and periodic when the energy is
real and open and nonperiodic when the energy is complex was made in
Ref.~\cite{R20} and studied in detail in Ref.~\cite{R33}. In these references it
is emphasized that the Bohr-Sommerfeld quantization condition 
\begin{equation}
\oint_C dx\,p=\left(n+\half\right)\pi
\label{e7}
\end{equation}
can only be applied if the classical orbits are closed. Thus, there is a deep
connection between real classical energies and the existence of associated real
quantum eigenvalues.

It was further argued in Ref.~\cite{R33} that the measurement of a quantum
energy is inherently imprecise because of the time-energy uncertainty principle
$\Delta E\,\Delta t\gtrsim\hbar/2$. Specifically, since there is not an infinite
amount of time in which to make a quantum energy measurement, we expect that the
uncertainty in the energy $\Delta E$ is nonzero. If we then suppose that this
uncertainty has an imaginary component, it follows that in the corresponding
classical theory, while the particle trajectories are almost periodic, the
orbits do not close exactly. The fact that the classical orbits with complex
energy are not closed means that in complex classical mechanics one can observe
tunneling-like phenomena that one normally expects to find only in quantum
systems.

\begin{figure}
\begin{center}
\includegraphics[scale=0.3, viewport=0 0 1000 617]{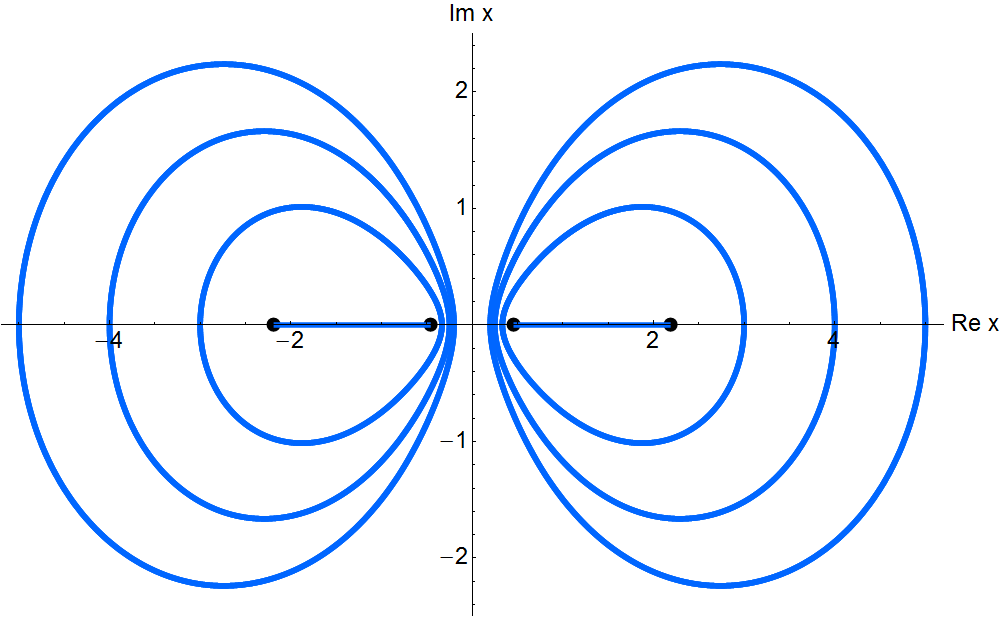}
\end{center}
\caption{Eight classical trajectories in the complex-$x$ plane representing a
particle of energy $E=-1$ in the potential $x^4-5x^2$. The turning points are
located at $x=\pm2.19$ and $x=\pm0.46$ and are indicated by dots. Because the
energy is real, the trajectories are all closed. The classical particle stays in
either the right-half or the left-half plane and cannot cross the imaginary
axis. Thus, when the energy is real, there is no effect analogous to tunneling.}
\label{F4}
\end{figure}

We illustrate such tunneling-like phenomena by considering a particle in a
quartic double-well potential $V(x)=x^4-5x^2$. Figure \ref{F4} shows eight
possible complex classical trajectories for a particle of {\it real} energy
$E=-1$. Each of these trajectories is closed and periodic. Observe that for
this energy the trajectories are localized either in the left well or the right
well and that no trajectory crosses from one side to the other side of the
imaginary axis.

\begin{figure}
\begin{center}
\includegraphics[scale=0.3, viewport=0 0 1000 976]{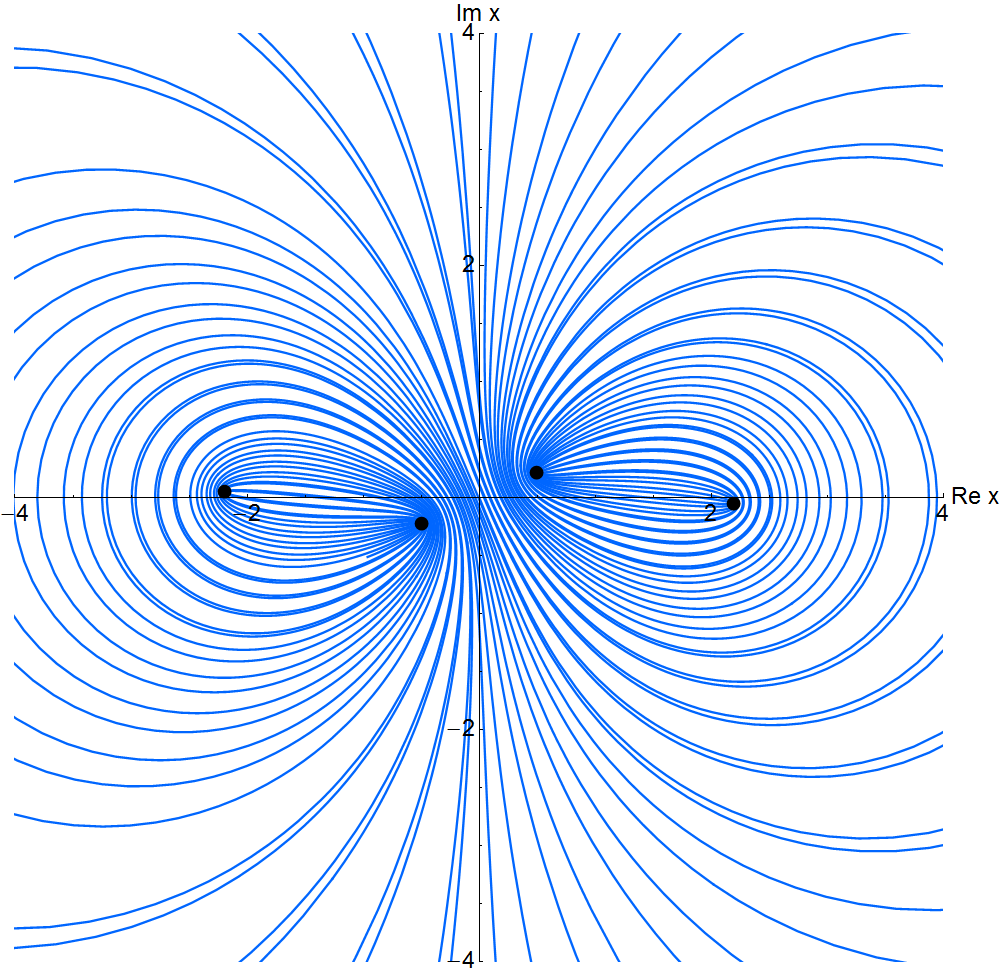}
\end{center}
\caption{Classical trajectory of a particle moving in the complex-$x$ plane
under the influence of a double-well $x^4-5x^2$ potential. The particle has
complex energy $E=-1-i$ and thus its trajectory does not close. The trajectory
spirals outward around one pair of turning points, crosses the imaginary axis,
and then spirals inward around the other pair of turning points. It then spirals
outward again, crosses the imaginary axis, and goes back to the original pair of
turning points. The particle repeats this behavior endlessly but at no point
does the trajectory cross itself. This classical-particle motion is analogous to
the behavior of a quantum particle that repeatedly tunnels between two
classically allowed regions. Here, the particle does not disappear into the
classically forbidden region during the tunneling process; rather, it moves
along a well-defined path in the complex-$x$ plane from one well to the other.}
\label{F5}
\end{figure}

What happens if we allow the classical energy to be complex? In this case the
classical trajectory is no longer closed. However, it does not spiral out to
infinity like the trajectory shown in Fig.~3. Rather, the trajectory in
Fig.~\ref{F5} unwinds around a pair of turning points for a characteristic
length of time and then crosses the imaginary axis. At this point the
trajectory does something remarkable: Rather than continuing its outward
journey, it spirals {\it inward} towards the other pair of turning points.
Then, never crossing itself, the trajectory turns outward again, and after
the same characteristic length of time, returns to the vicinity of the first
pair of turning points. This oscillatory behavior, which shares the qualitative
characteristics of strange attractors, continues forever but the trajectory
never crosses itself. As in the case of quantum tunneling, the particle spends 
a long time in proximity to a given pair of turning points before crossing
the imaginary axis to the other pair of turning points. On average, the
classical particle spends equal amounts of time on either side of the imaginary
axis. Interestingly, we find that as the imaginary part of the classical
energy increases, the characteristic ``tunneling'' time decreases in inverse
proportion, just as one would expect of a quantum particle.

\begin{figure}
\begin{center}
\includegraphics[scale=0.23, viewport=0 0 1000 786]{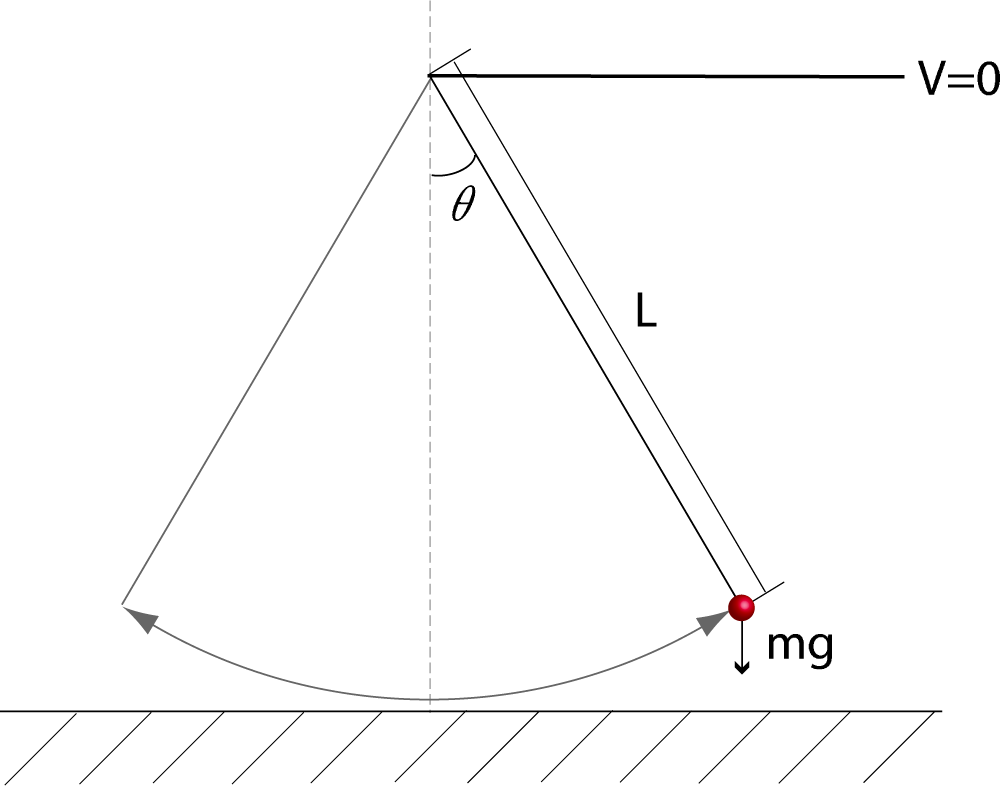}
\end{center}
\caption{Configuration of a simple pendulum of mass $m$ in a uniform
gravitational field of strength $g$. The length of the string is $L$. The
pendulum swings through an angle $\theta$. We define the potential energy to be
$0$ at the height of the pivot.
\label{F6}}
\end{figure}

Having described the tunneling-like behavior of a classical particle having
complex energy in a double well, we examine the case of such a particle in a
periodic potential. Physically, this corresponds to a classical particle in
a crystal lattice. A simple physical system that has a periodic potential
consists of a simple pendulum in a uniform gravitational field \cite{R34}.
Consider a pendulum consisting of a bob of mass $m$ and a string of length $L$
in a uniform gravitational field of magnitude $g$ (see Fig.~\ref{F6}). The
gravitational potential energy of the system is defined to be zero at the height
of the pivot point of the string. The pendulum bob swings through an angle
$\theta$. Therefore, the horizontal and vertical cartesian coordinates $X$ and
$Y$ are $X=L\sin\theta$ and $Y=-L\cos\theta$, which gives velocities $\dot{X}=L
\dot{\theta}\sin\theta$ and $\dot{Y}=-L\dot{\theta}\cos\theta$. The potential
and kinetic energies are $V=-mgL\cos\theta$ and $T=\half m(\dot{X}^2+\dot{Y}^2)=
\half mL^2\dot{\theta}^2$. The Hamiltonian $H=T+V$ for the pendulum is therefore
\begin{equation}
H=\half mL^2\dot{\theta}^2-mgL\cos\theta.
\label{e8}
\end{equation}

Without loss of generality we set $m=1$, $g=1$, and $L=1$ and then make the
change of variable $\theta\rightarrow x$ to get
\begin{equation}
H=\half p^2-\cos x,
\label{e9}
\end{equation}
where $p=\dot{x}$. The classical equations of motion for this Hamiltonian are
\begin{equation}
\dot{x}=\frac{\partial H}{\partial p}=p,\qquad\dot{p}=-\frac{\partial H}{
\partial x}=-\sin x.
\label{e10}
\end{equation}
The Hamiltonian $H$ for this system is a constant of the motion and thus the
energy $E$ is a time-independent quantity.

\begin{figure}
\begin{center}
\includegraphics[scale=0.36, viewport=0 0 1000 434]{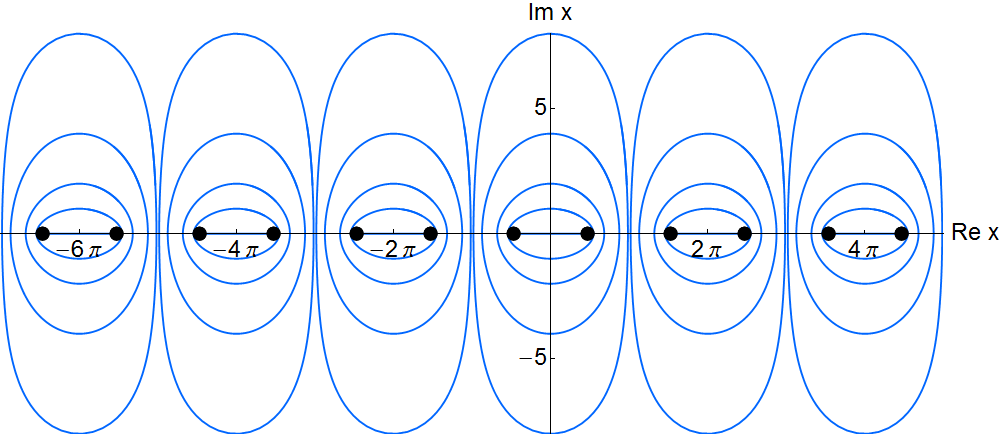}
\end{center}
\caption{Classical trajectories in the complex-$x$ plane for a particle of
energy $E=-0.09754$ in a $-\cos x$ potential. The motion is periodic and the
particle remains confined to a cell of width $2\pi$. Five trajectories are shown
for each cell. The trajectories shown here are the periodic analogs of the
trajectories shown in Figs.~\ref{F1} and \ref{F2}.}
\label{F7}
\end{figure}

If we take the energy $E$ to be real, we find that the classical trajectories
are confined to cells of horizontal width $2\pi$, as shown in Fig.~\ref{F7}.
This is the periodic analog of Figs.~\ref{F1} and \ref{F2}.

\begin{figure}
\begin{center}
\includegraphics[scale=0.50, viewport=0 0 766 493]{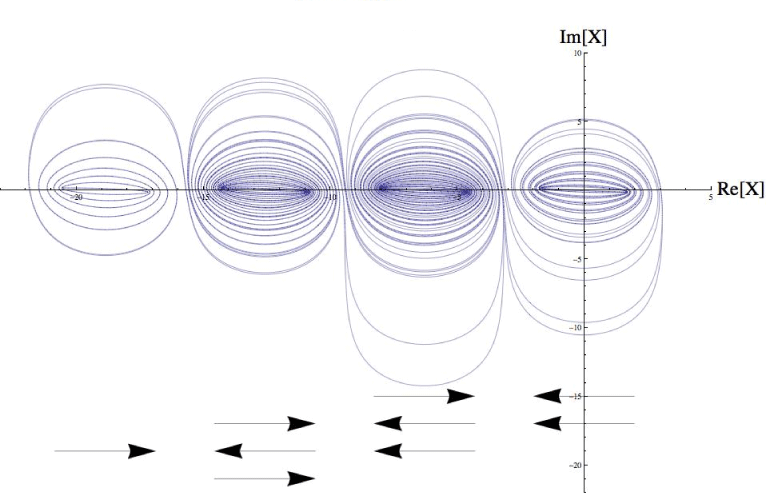}
\end{center}
\caption{A tunneling trajectory for the Hamiltonian (\ref{e9}) with $E=0.1-0.15
i$. The classical particle hops from well to well in a random-walk fashion. The
particle starts at the origin and then hops left, right, left, left, right,
left, left, right, right. This is the sort of behavior normally associated with
a particle in a crystal at an energy that is not in a conduction band. At the
end of this simulation the particle is situated to the left of its initial
position. The trajectory never crosses itself.}
\label{F8}
\end{figure}

If the energy of the classical particle in a periodic potential is taken to be
complex, the particle begins to hop from well to well in analogy to the behavior
of the particle in Fig.~\ref{F5}. This hopping behavior is displayed in
Fig.~\ref{F8}.

\begin{figure}
\begin{center}
\includegraphics[scale=0.36, viewport=0 0 1217 777]{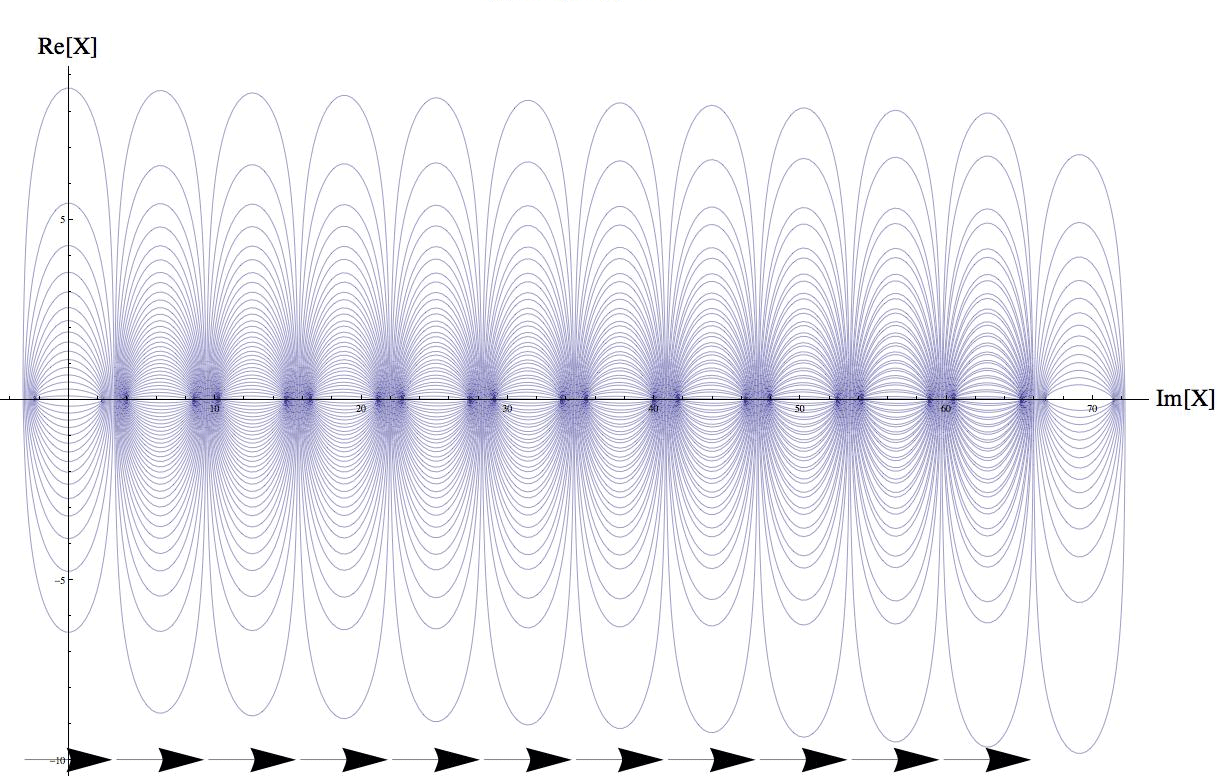}
\end{center}
\caption{A classical particle exhibiting a behavior analogous to that of a
quantum particle in a conduction band that is undergoing resonant tunneling.
Unlike the particle in Fig.~\ref{F8}, this classical classical particle tunnels
in one direction only and drifts at a constant average velocity through the
potential.}
\label{F9}
\end{figure}

The most interesting analogy between quantum mechanics and complex classical
mechanics is established by showing that there exist narrow conduction bands in
the periodic potential for which the quantum particle exhibits resonant
tunneling and the complex classical particle exhibits unidirectional hopping
\cite{R33,R35}. This qualitative behavior is illustrated in Fig.~\ref{F9}.

\begin{figure}
\includegraphics[scale=0.45, viewport=0 0 1024 668]{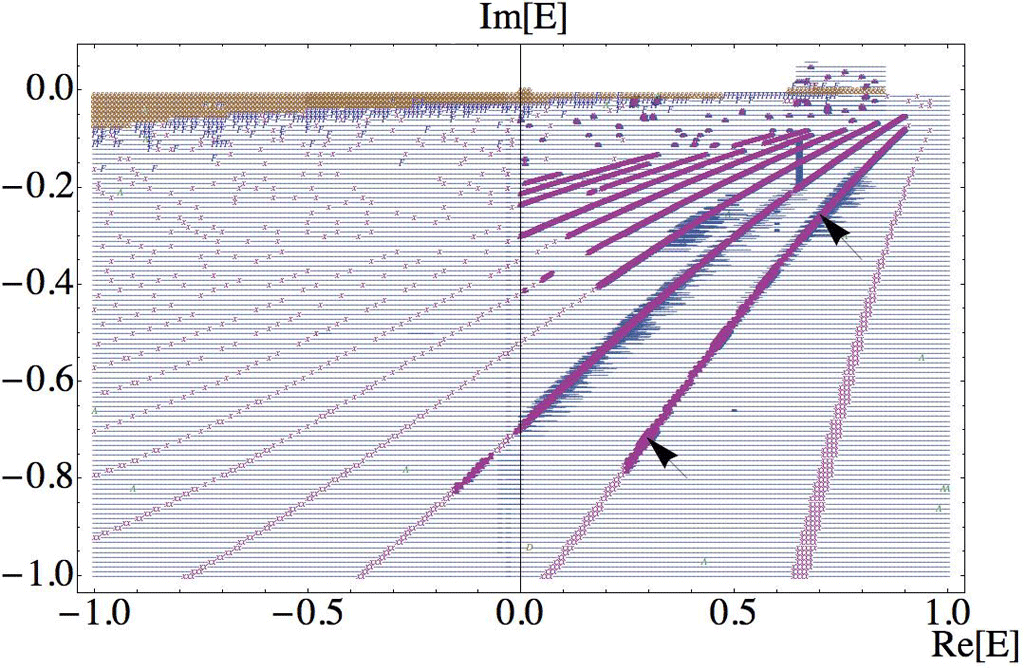}
\caption{Complex-energy plane showing those energies that lead to tunneling
(hopping) behavior and those energies that give rise to conduction. Hopping
behavior is indicated by a hyphen - and conduction is indicated by an X. The
symbol \& indicates that no tunneling takes place; tunneling does not occur for
energies whose imaginary part is close to 0. In some regions of the energy plane
we have done very intensive studies and the X's and -'s are densely packed. This
picture suggests the features of band theory: If the imaginary part of the
energy is taken to be $-0.9$, then as the real part of the energy increases from
$-1$ to $+1$, five narrow conduction bands are encountered. These bands are
located near ${\rm Re}\,E=-0.95,\,-0.7,\,-0.25,\,0.15,\,0.7$. This picture is
symmetric about ${\rm Im}\,E=0$ and the bands get thicker as $|{\rm Im}\,E|$
increases. A total of 68689 points were classified to make this plot. In most
places the resolution (distance between points) is $0.01$, but in several
regions the distance between points is shortened to $0.001$. The regions
indicated by arrows are blown up in Fig.~\ref{F11}.}
\label{F10}
\end{figure}

\begin{figure}
\begin{center}
\includegraphics[scale=0.21, viewport=0 0 2122 595]{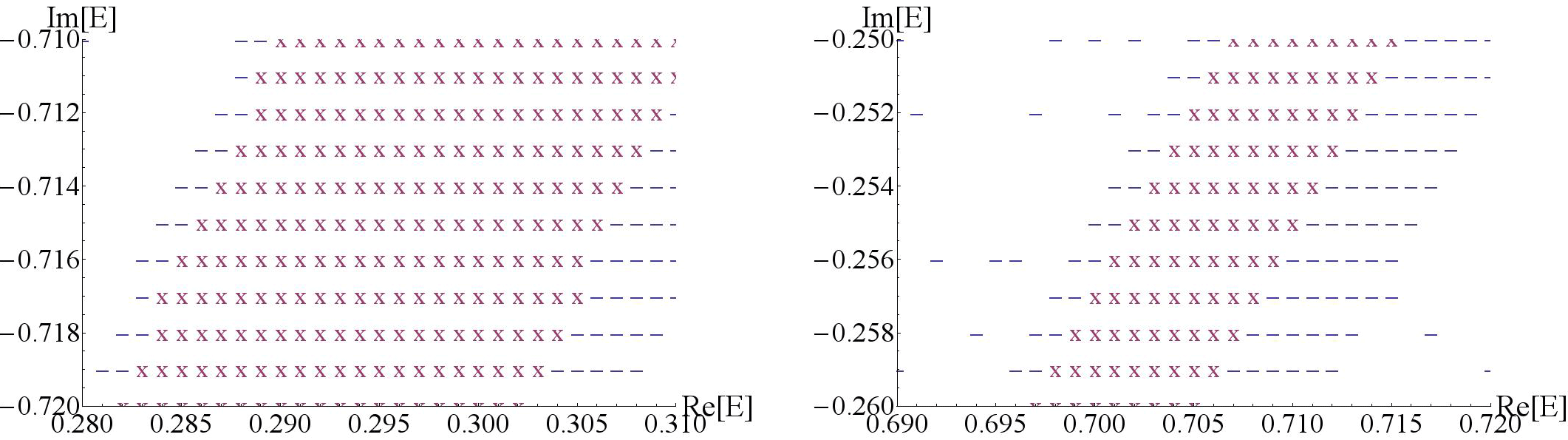}
\end{center}
\caption{Detailed portions of the complex-energy plane shown in Fig.~\ref{F10}
containing a conduction band. Note that the edge of the conduction band, where
tunneling (hopping) behavior changes over to conducting behavior, is very
sharp.}
\label{F11}
\end{figure}

A detailed numerical analysis shows that the classical conduction bands have a
narrow but finite width (see Fig.~\ref{F10}). Two magnified portions of the
conduction bands in Fig.~\ref{F10} are shown in Fig.~\ref{F11}. These
magnifications show that the edges of the conduction bands are sharply defined. 

\section{Classical Particle in a Complex Elliptic Potential}
\label{s3}

Having reviewed in Sec.~\ref{s2} the behavior of complex classical trajectories
for trigonometric potentials, in this section we give a brief glimpse of the
rich and interesting behavior of classical particles moving in elliptic
potentials. Elliptic potentials are natural doubly-periodic generalizations
of trigonometric potentials. 

\begin{figure}
\begin{center}
\includegraphics[scale=0.45, viewport=0 0 1000 522]{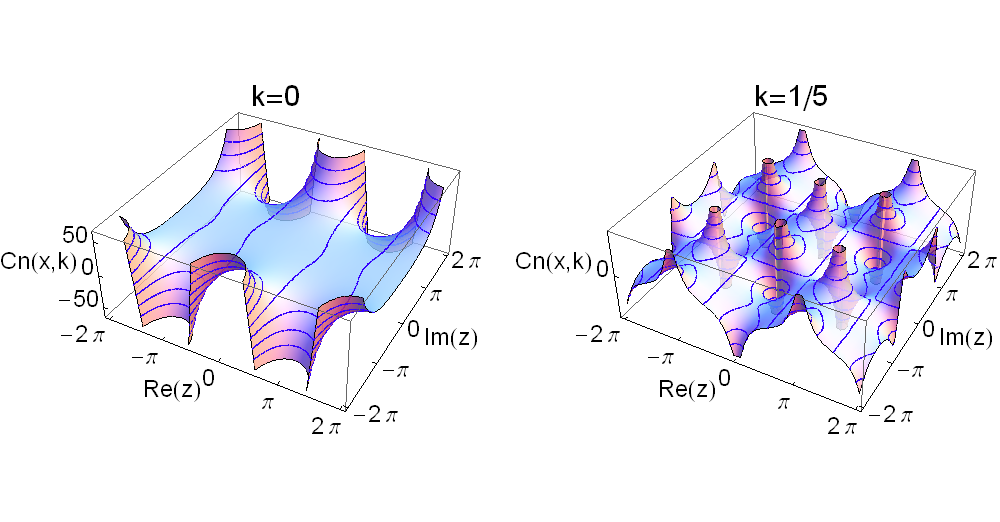}
\end{center}
\caption{Real part of the cnoidal elliptic function ${\rm Cn}(x,k)$ in the
complex-$x$ plane for two values of $k$, $k=0$ and $k=1/5$. When $k=0$, this
cnoidal function reduces to the trigonometric function $\cos x$, and this
function grows exponentially in the imaginary-$x$ direction. When $k>0$, the
cnoidal function is doubly periodic; that is, periodic in the real-$x$ and in
the imaginary-$x$ directions. While $\cos x$ is entire, the cnoidal functions 
for $k\neq0$ are meromorphic and have periodic double poles.}
\label{F12}
\end{figure}

The Hamiltonian that we have chosen to study is a simple extension of that
in (\ref{e9}):
\begin{equation}
H=\half p^2-{\rm Cn}(x,k),
\label{e11}
\end{equation}
where ${\rm Cn}(x,k)$ is a {\it cnoidal} function \cite{R36,R37}. When the
parameter $k=0$, the cnoidal function reduces to the singly periodic function
$\cos x$ and when $k=1$ the cnoidal function becomes ${\rm tanh}\,x$. When $0<
k<1$, the cnoidal function is periodic in both the real and imaginary directions
and it is meromorphic (analytic in the finite-$x$ plane except for pole
singularities) and has infinitely many double poles. The real part of the
cnoidal potential ${\rm Cn}(x,k)$ is plotted in Fig.~\ref{F12}.

\begin{figure}
\begin{center}
\includegraphics[scale=0.26, viewport=0 0 1000 1426]{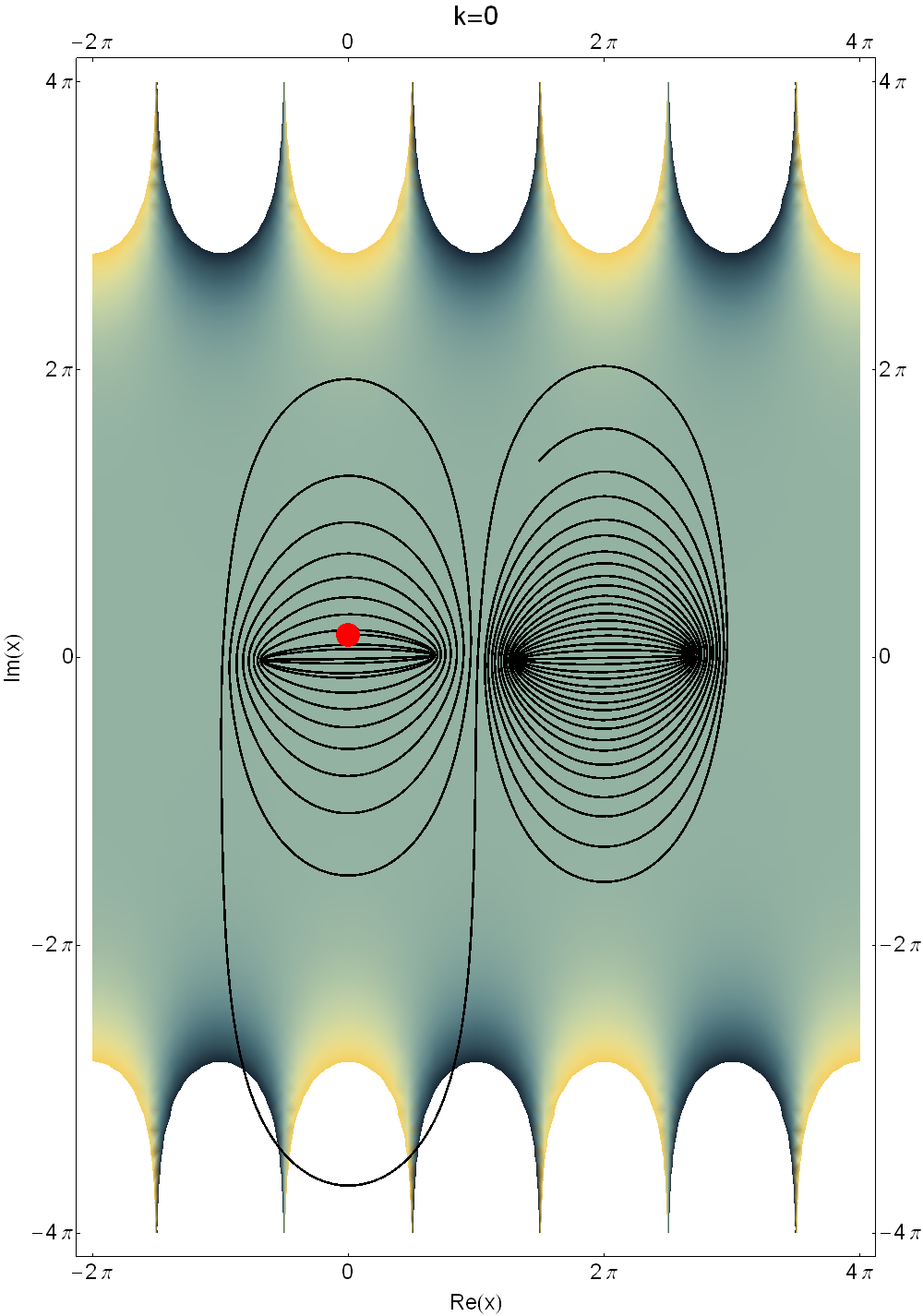}
\end{center}
\caption{Complex classical trajectory for a particle of energy $E=0.5+0.05i$ in
a cosine potential (a cnoidal potential with $k=0$). The trajectory begins at
the red dot at $x=0.5i$ on the left side of the figure and spirals outward
around the left pair of turning points. The trajectory then "tunnels" to the
right and spirals inward and then outward around the right pair of turning
points. In the background is a plot of the real part of the cosine potential,
which is shown in detail in Fig.~\ref{F12}.}
\label{F13}
\end{figure}

\begin{figure}
\begin{center}
\includegraphics[scale=0.45, viewport=0 0 1000 483]{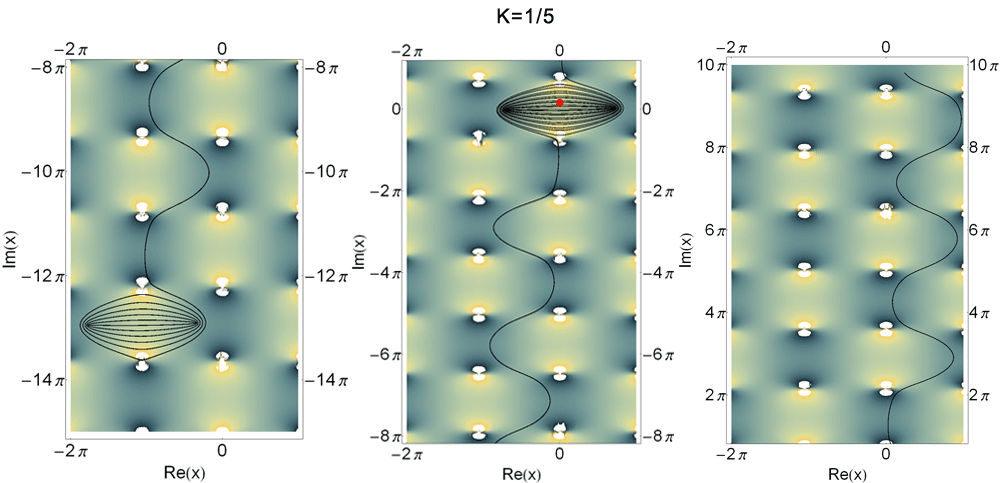}
\end{center}
\caption{Complex classical trajectory for a particle of energy $E=0.5+0.05i$ in
a cnoidal potential with $k=1/5$. The trajectory begins at the red dot at
$x=0.5i$ near the top of the central pane. The trajectory spirals outward around
the two turning points and, when it gets very close to a pole, it suddenly
begins to travel downward in a wavy fashion in the negative-imaginary direction.
The trajectory bobs and weaves past many double poles and continues on in the
left pane of the figure. It eventually gets very close to a double pole, gets
trapped, and spirals inward towards a pair of turning points. After spiraling
inwards, it then spirals outwards (never crossing itself) and goes upward along
a path extremely close to the downward wavy path. It is then recaptured by the
original pair of turning points in the central pane. After spiraling inwards and
outwards once more it now escapes and travels {\it upward}, bobbing and weaving
along a wavy path in the right pane of the figure. Evidently, the particle
trajectory strongly prefers to move vertically upward and downward, and not
horizontally. The vertical motion distinguishes cnoidal trajectories from those
in Figs.~\ref{F8}, \ref{F9}, and \ref{F13} associated with the cosine
potential.}
\label{F14}
\end{figure}

The classical particle trajectories satisfy Hamilton's equations
\begin{eqnarray}
\dot{x}&=&\frac{\partial H}{\partial p}=p,\nonumber\\
\dot{p}&=&-\frac{\partial H}{\partial x}={\rm Sn}(x,k){\rm Dn}(x,k).
\label{e12}
\end{eqnarray}
The trajectories for $k>0$ are remarkable in that the classical particles seem
to prefer to move vertically rather than horizontally. In Fig.~\ref{F13} a
trajectory, similar to that in Fig.~\ref{F8}, is shown for the case $k=0$.
This trajectory is superimposed on a plot of the real part of the cosine
potential. The particle oscillates horizontally. In Fig.~\ref{F14} a complex
trajectory for the case $k=1/5$ is shown. This more exotic path escapes from the
initial pair of turning points, and rather than "tunneling" to a horizontally
adjacent pair of turning points, it travels downward. The ensuing wavy vertical
motion passes close to many poles before the particle gets captured by another
pair of turning points. The particle winds inwards and outwards around these
turning points and eventually returns to the original pair of turning points.
After escaping from these turning points again, the particle now moves in the 
positive-imaginary direction.

It is clear that classical trajectories associated with doubly periodic
potentials have an immensely interesting structure and should be investigated in
much greater detail to determine if there is a behavior analogous to band
structure shown in Figs.~\ref{F10} and \ref{F11}.

\section{Summary and Discussion}
\label{s4}

The relationship between quantum mechanics and classical mechanics is subtle.
Quantum mechanics is essentially wavelike; probability amplitudes are described
by a wave equation and physical observations involve such wavelike phenomena as
interference patterns and nodes. In contrast, classical mechanics describes the
motion of particles and exhibits none of these wavelike features. Nevertheless,
there is a deep connection between quantum mechanics and complex classical
mechanics. In the complex domain the classical trajectories exhibit a
remarkable behavior that is analogous to quantum tunneling.

Periodic potentials exhibit a surprising and intricate feature that closely
resembles quantum band structure. It is especially noteworthy that the classical
bands, just like the quantum bands, have finite width.

Our early work on singly periodic potentials strongly suggests that further
detailed analysis should be done on doubly periodic potentials. Doubly periodic
potentials are particularly interesting because they have singularities. Two
important and so far unanswered questions are as follows: (i) Does a complex
classical particle in a doubly periodic potential undergo a random walk in two
dimensions and eventually visit all lattice sites? (ii) Are there special bands
of energy for which the classical particle no longer undergoes random hopping
behavior and begins to drift in one direction through the lattice?

\begin{acknowledgments}
CMB is grateful to Imperial College for its hospitality and to the
U.S.~Department of Energy for financial support. DWH thanks Symplectic
Ltd.~for financial support. Mathematica was used to generate the figures in
this paper.
\end{acknowledgments}

\end{document}